\documentclass[%
 reprint,
 superscriptaddress,
 showpacs,preprintnumbers,
 amsmath,amssymb,
 aps,
 pre,
 floatfix,
]{revtex4-1}

\usepackage{graphicx}
\usepackage{dcolumn}
\usepackage{bm}
\usepackage{hyperref}

\newcolumntype{C}[1]{>{\centering\let\newline\\\arraybackslash\hspace{0pt}}m{#1}}

\begin{document}

\preprint{}

\title{Stability of the $\omega$ structure of transition elements}

\author{Yuji Ikeda}
\email{ikeda.yuji.6m@kyoto-u.ac.jp}
\affiliation{Center for Elements Strategy Initiative for Structure Materials (ESISM), Kyoto University, Kyoto 606-8501, Japan}

\author{Isao Tanaka}
\email{tanaka@cms.mtl.kyoto-u.ac.jp}
\affiliation{Center for Elements Strategy Initiative for Structure Materials (ESISM), Kyoto University, Kyoto 606-8501, Japan}
\affiliation{Department of Materials Science and Engineering, Kyoto University, Kyoto 606-8501, Japan}
\affiliation{Center for Materials Research by Information Integration, National Institute for Materials Science (NIMS), Tsukuba 305-0047, Japan}
\affiliation{Nanostructures Research Laboratory, Japan Fine Ceramics Center, Nagoya 456-8587, Japan}


\date{\today}

\begin{abstract}
Properties of the $\omega$ structure are investigated for 27 transition elements
from the viewpoints of thermodynamical and mechanical stability
based on first-principles calculations.
The thermodynamical stability of the $\omega$ structure is compared with
those for the body-centered cubic (BCC),
face-centered cubic (FCC),
and hexagonal close-packed (HCP) structures.
Similarly to the case of those popular crystal structures,
the occupation number for $d$ orbitals is found to roughly determine
relative energy and volume of the nonmagnetic (NM) $\omega$ structure.
For the group 4 elements (Ti, Zr, and Hf),
the $\omega$ structure is almost the lowest in energy
among the investigated crystal structures
and is also mechanically stable.
The $\omega$ structure of the group 7 elements (Mn, Tc, and Re) is also
mechanically stable.
The $\omega$ Fe is found to exhibit a complicated magnetic state
that is different from the ferromagnetic (FM) and NM ones.
This magnetic state is the most favorable among the investigated magnetic states.
The $\omega$ Fe in this magnetic state is also mechanically stable.
Energies of binary alloys composed of the elements in the group 4
and those in the groups 5 and 6 are estimated by linear interpolation,
and most of the alloys show concentration ranges
where the $\omega$ structure is the lowest in energy
among the investigated crystal structures.
\end{abstract}

\pacs{
71.20.Be, 
61.50.-f, 
71.15.Mb, 
63.20.-e, 
63.20.dk  
}

\maketitle


\section{INTRODUCTION}
\label{sec:introduction}

The $\omega$ structure is included in the hexagonal crystal system
with the space group of $P6/mmm$ (No. 191).
It was first reported for Ti-Cr alloys,
and its relation to the brittleness of the alloys was discussed
\cite{Frost1954}.
The $\omega$ structure is observed for elemental
Ti \cite{Jamieson1963},
Zr \cite{Jamieson1963},
and
Hf~\cite{Xia1990, Pandey2014}
under high pressure.
Ti and Zr can hold the $\omega$ structure also
after removing the pressure
\cite{Jamieson1963}.
As well as the pure elements,
the $\omega$ structure is observed
for alloys based on the group 4 elements, namely
Ti-%
\cite{Frost1954, Silcock1958, Sass1969, Hickman1969_TMSAIME, Cui2009, Wu2014, Ping2006},
Zr-%
\cite{Hatt1957, Hatt1960, Sass1969},
and
Hf-%
\cite{Jackson1970}
based alloys.
In addition,
it has been reported that
the $\omega$ structure is formed in elemental Ta and Ta-W alloys
by applying shock pressure
\cite{Hsiung2000}
and in elemental Mo after high-pressure torsion
\cite{Cheng2013}.
Several experimental reports have recently claimed that
the $\omega$ structure can be found also in steels,
i.e., Fe-C-based alloys
\cite{Ping2013, Liu2015}.
Structures based on the $\omega$ lattice,
where constituent elements occupy the same atomic sites as the $\omega$ structure
with atomic orderings,
have also been observed in experiments for alloys such as
Cu-Zn \cite{Prasetyo1976_CuZn},
Cu-Mn-Al \cite{Prasetyo1976_CuMnAl},
Ni-Al \cite{Georgopoulos1981},
Fe-Ni-Co-Mo \cite{Yedneral1972},
Fe-Mn-Co-Mo \cite{Lecomte1985},
and Fe-Ni-Mo \cite{Djega-Mariadassou1995}
alloys.

The $\omega$ structure can be obtained
via collective motion of atoms
from the body-centered cubic (BCC)
\cite{Sikka1982},
hexagonal-close packed (HCP)
\cite{Silcock1958, Usikov1973, Trinkle2003},
and face-centered cubic (FCC)
\cite{Togo2013}
structures.
It is, therefore, suggested that
the $\omega$ structure can be the transition state of a transformation pathway
between these popular crystal structures.
\citeauthor{Togo2013} have actually revealed that
the $\omega$ structure for Cu can be the transition state
of a BCC-FCC transformation pathway
based on a systematic search algorithm for transformation pathways
\cite{Togo2013}.
\citeauthor{Ikeda2014} have also pointed out that
the $\omega$ structure for Fe acts
as the transition state of the pressure-induced phase transition
between the high-temperature paramagnetic (PM) BCC and PM FCC structures
\cite{Ikeda2014}.
These results imply that
the $\omega$ structure is important
not only for the metals and the alloys that form the $\omega$ structure
but also for general metallic systems to understand mechanisms of phase transitions.
However, systematic knowledge of the $\omega$ structure for metallic systems is still missing.

It is interesting that the elements in the groups 5 and 6 such as V, Cr, Nb, and Mo are
included in most of the alloys based on the group 4 elements that form the $\omega$ structure.
For these alloys,
the $\omega$ structure is observed in a concentration range
where the group 4 elements are rich.
These experimental facts imply that interactions between the group 4 elements
and those in the groups 5 and 6 have essential roles to form the $\omega$ structure.
To our best knowledge, however,
no detailed and systematic investigations have been accomplished into this issue.

In this study,
properties of the $\omega$ structure are
systematically investigated for 27 transition elements
from the viewpoints of thermodynamical and mechanical stability
based on first-principles calculations.
The thermodynamical stability of the $\omega$ structure is compared with
those of the BCC, FCC, and HCP structures.
The mechanical stability of the $\omega$ structure is investigated
in terms of phonon frequencies.
We also investigate thermodynamical stability of the $\omega$ structure
for binary alloys composed of the transition elements.

\section{COMPUTATIONAL DETAILS}

\subsection{$\omega$ structure}

\begin{figure}[tbp]
\begin{center}
\includegraphics[width=\linewidth]{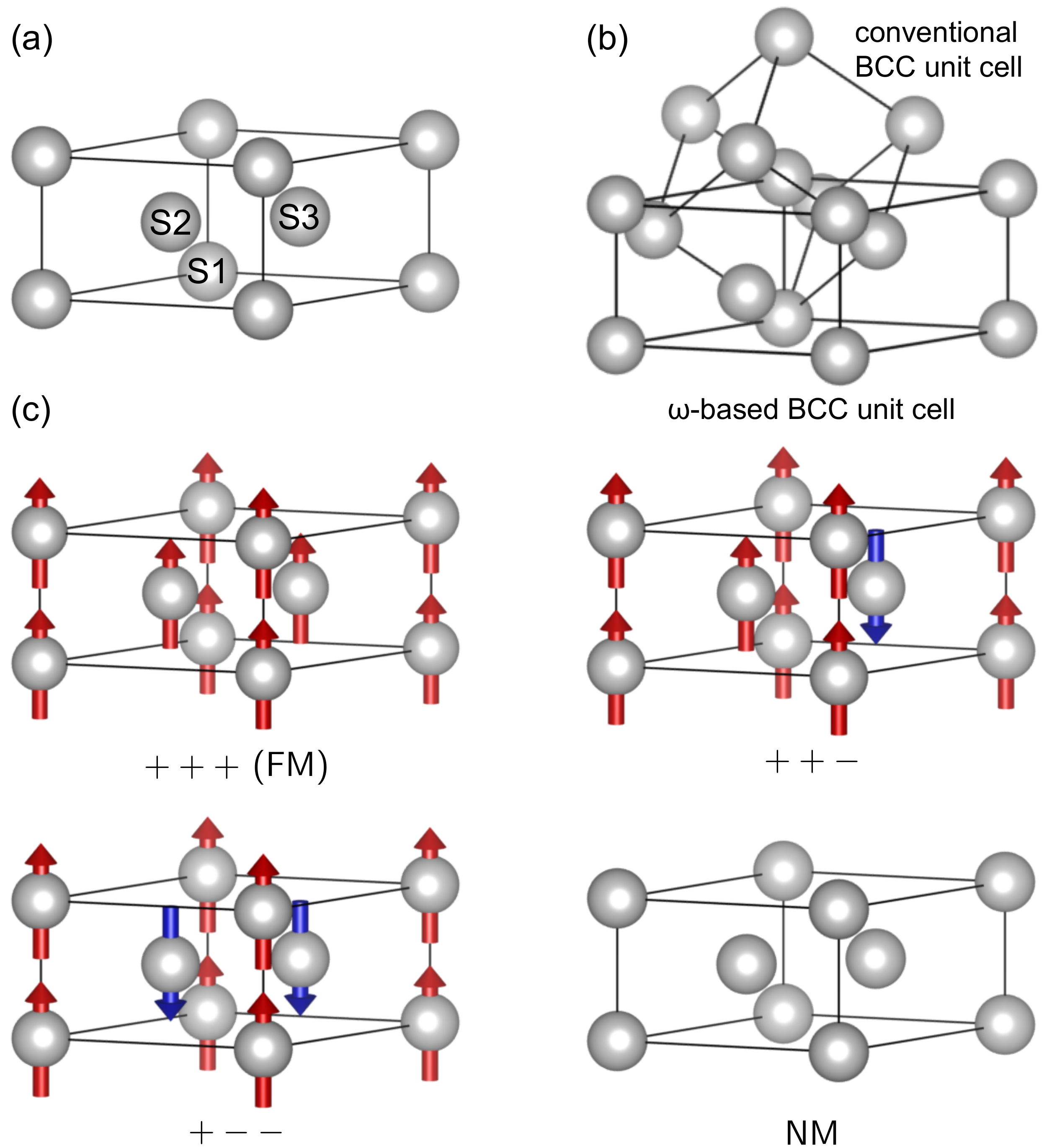}
\end{center}
\caption{
(Color online)
(a) Primitive unit cell of the $\omega$ structure.
Gray spheres represent atoms.
The labels S1, S2, and S3 specify the atoms whose positions are described
in the main text.
(b) Geometrical relation of the $\omega$-based BCC unit cell
to the conventional BCC unit cell.
(c) Magnetic states of the $\omega$ structure investigated in this study.
Up (red) and down (blue) arrows indicate spin-up and spin-down magnetic moments,
respectively.
Visualization is performed using the \textsc{VESTA} code
\cite{Momma2011}.
\label{fig:magnetic_configurations}
}
\end{figure}

Figure~\ref{fig:magnetic_configurations}(a) shows the primitive unit cell
of the $\omega$ structure without considering magnetic configurations.
The basis of lattice vectors for the $\omega$ structure
$\mathbf{a}_{1}$,
$\mathbf{a}_{2}$,
and
$\mathbf{a}_{3}$
can be written as,
\begin{align}
\begin{pmatrix}
\mathbf{a}_{1} & \mathbf{a}_{2} & \mathbf{a}_{3}
\end{pmatrix}
&=
\begin{pmatrix}
a_{\omega}/2 & a_{\omega}/2 & 0 \\
-\sqrt{3}a_{\omega}/2 & \sqrt{3}a_{\omega}/2 & 0 \\
0 & 0 & c_{\omega}
\end{pmatrix},
\end{align}
where $a_{\omega}$ and $c_{\omega}$ are lattice constants of the $\omega$ structure.
The $\omega$ structure has three atoms,
referred to as S1, S2, and S3 hereafter,
inside the primitive unit cell.
The atomic positions of S1, S2, and S3 are
$(0, 0, 0)$,
$(2/3, 1/3, 1/2)$,
and
$(1/3, 2/3, 1/2)$
in fractional coordinates,
respectively.
The sites S2 and S3 are crystallographycally equivalent
without considering magnetic configurations.
The Wyckoff positions are $1a$ for the atom S1 and $2d$ for the atoms S2 and S3.

The BCC structure is related to the $\omega$ structure
by the orientation relation of
$[0001]_\omega || \langle 111 \rangle_{\textrm{BCC}}$ and
$\{11\bar{2}0\}_\omega || \{1\bar{1}0\}_{\textrm{BCC}}$.
The BCC structure can be obtained
from the primitive $\omega$ unit cell
by moving the atoms S2 and S3 to $(2/3, 1/3, 1/3)$ and $(1/3, 2/3, 2/3)$
in fractional coordinates, respectively,
and by modifying the lattice constants of the $\omega$ structure as,
\begin{align}
a_{\omega} &= \sqrt{2} a_{\textrm{BCC}},
\label{eq:a_omega}
\\
c_{\omega} &= \sqrt{3} a_{\textrm{BCC}} / 2,
\label{eq:c_omega}
\end{align}
where $a_{\textrm{BCC}}$ is the lattice constant of the BCC structure.
Figure~\ref{fig:magnetic_configurations}(b) shows the geometrical relation
between the conventional and the ``$\omega$-based'' unit cell of the BCC structure.
From the viewpoint of the $\omega$ structure,
it is ``coherent'' with the BCC
when the $\omega$ structure has the lattice constants
that satisfy Eqs.~(\ref{eq:a_omega}) and (\ref{eq:c_omega}).
In other words,
if the lattice constants of the $\omega$ and the BCC structures
exactly satisfy Eqs.~(\ref{eq:a_omega}) and (\ref{eq:c_omega}),
the $\omega$ structure can be obtained from the BCC
by moving atoms without lattice deformation.

As well as the $\omega$ structure,
the BCC, FCC, and HCP structures were also investigated for comparison.
Four possible magnetic states,
including the ferromagnetic (FM) and the nonmagnetic (NM) states,
were considered for the $\omega$ structure.
Figure~\ref{fig:magnetic_configurations}(c) shows the considered magnetic
states for the $\omega$ structure.
For the BCC, FCC, and HCP structures,
only the FM and the NM states were considered.

\subsection{Electronic structures and phonons}

\begin{table}[tbp]
\begin{center}
\caption{
Meshes per unit cell to sample Brillouin zones.
\label{tb:meshes}
}
\begin{ruledtabular}
\begin{tabular}{ccc}
&
Number of atoms per unit cell &
Mesh \\
\hline
BCC      &        2 & $16 \times 16 \times 16$\\
FCC      &        4 & $12 \times 12 \times 12$\\
HCP      &        2 & $18 \times 18 \times 12$\\
$\omega$ &        3 & $12 \times 12 \times 18$\\
\end{tabular}
\end{ruledtabular}
\end{center}
\end{table}

The plane-wave basis projector augmented wave
method
\cite{Blochl1994} 
was employed in the framework of
density-functional theory
\cite{Hohenberg1964,Kohn1965}
within the generalized gradient approximation
(GGA)
in the Perdew-Burke-Ernzerhof
(PBE)
form
\cite{Perdew1996} 
as implemented in the \textsc{VASP} code
\cite{
Kresse1995, 
Kresse1996, 
Kresse1999}.
A plane-wave energy cutoff of 400~eV was used.
%
The Brillouin zones were sampled by $\Gamma$ centered meshes
according to crystal structures as shown in Table~\ref{tb:meshes},
and the Methfessel-Paxton scheme~\cite{Methfessel1989} with a smearing width of 0.4~eV
was employed.
Total energies were minimized until the energy convergences to be
less than 10$^{-8}$~eV.
Lattice parameters were optimized
under zero external stress.
%
Magnetic moments on atoms were determined from the electron density
in corresponding Voronoi cells.

Phonon frequencies of the $\omega$ structures were calculated based on the harmonic approximation
for a lattice Hamiltonian using the finite-displacement method.
Atomic displacements of 0.01~{\AA} for
the $2 \times 2 \times 4$ supercell of the $\omega$ unit cell (including 48 atoms)
were used to calculate the second-order force constants.
The \textsc{PHONOPY} code~\cite{Togo2015, Togo2008} was used
for these phonon calculations.

\section{RESULTS AND DISCUSSION}
\label{sec:results}

\subsection{Energetics for the $\omega$ structure of transition elements}

\begin{table}[tbp]
\begin{center}
\caption{
Obtained magnetic states with nonzero magnetic moments.
\label{tb:magnetic_states}
}
\begin{ruledtabular}
\begin{tabular}{lll}
 &
\multicolumn{1}{c}{Magnetic state} &
\multicolumn{1}{c}{Element} \\
\hline
BCC      & FM    & Mn, Fe, Co, Ni, Rh, Ir\\
\\
FCC      & FM    & Fe, Co, Ni\\
\\
HCP      & FM    & Co, Ni\\
\\
$\omega$ & FM    & Fe, Co, Ni\\
         & $++-$ & Fe\\
         & $+--$ & Fe\\
\end{tabular}
\end{ruledtabular}
\end{center}
\end{table}

Table~\ref{tb:magnetic_states} summarizes the magnetic states
with nonzero magnetic moments found in the present calculations.
Most of the nonzero magnetic moments are found for the $3d$ transition elements.
The FM $\omega$ structure is obtained for Fe, Co, and Ni.
The $\omega$ Fe has also the $++-$ and the $+--$ magnetic states,
which are described in Fig.~\ref{fig:magnetic_configurations}(c).
Magnetic moments converge to zero for the rest of the elements
even for spin-unrestricted calculations.


\begin{table}[tbp]
\begin{center}
\caption{
Calculated energies,
lattice constants, and volumes
of the NM $\omega$ structures for the transition elements.
The energies are relative to that of the NM FCC structure.
\label{tb:energies_elements_nm}
}
\begin{ruledtabular}
\begin{tabular}{ldddd}
 &
\multicolumn{1}{c}{Relative energy} &
\multicolumn{1}{c}{$a_\omega$} &
\multicolumn{1}{c}{$c_\omega$} &
\multicolumn{1}{c}{Volume} \\
 &
\multicolumn{1}{c}{(meV/atom)} &
\multicolumn{1}{c}{(\AA)} &
\multicolumn{1}{c}{(\AA)} &
\multicolumn{1}{c}{(\AA$^{3}$/atom)} \\
\hline
Sc    &    8  & 5.098  & 3.202  & 24.02 \\
Ti    &  -70  & 4.542  & 2.824  & 16.82 \\
V     & -142  & 4.448  & 2.341  & 13.37 \\
Cr    &  -66  & 4.229  & 2.262  & 11.68 \\
Mn    &    6  & 3.887  & 2.448  & 10.68 \\
Fe    &  91   & 3.850  & 2.418  & 10.35 \\
Co    &  166  & 3.899  & 2.390  & 10.49 \\
Ni    &  111  & 3.987  & 2.397  & 11.00 \\
Cu    &  77   & 4.126  & 2.472  & 12.15 \\
\\
Y     &   37  & 5.638  & 3.527  & 32.36 \\
Zr    &  -40  & 5.036  & 3.149  & 23.06 \\
Nb    & -125  & 4.876  & 2.678  & 18.38 \\
Mo    &  -35  & 4.668  & 2.538  & 15.97 \\
Tc    &   -7  & 4.291  & 2.716  & 14.44 \\
Ru    &  137  & 4.252  & 2.680  & 13.99 \\
Rh    &  223  & 4.329  & 2.655  & 14.37 \\
Pd    &  107  & 4.493  & 2.676  & 15.59 \\
Ag    &   65  & 4.721  & 2.820  & 18.15 \\
\\
Lu    &   49  & 5.433  & 3.413  & 29.08 \\
Hf    &  -34  & 4.963  & 3.096  & 22.01 \\
Ta    &  -28  & 4.853  & 2.709  & 18.42 \\
W     &   51  & 4.678  & 2.590  & 16.36 \\
Re    &  106  & 4.375  & 2.744  & 15.16 \\
Os    &  209  & 4.320  & 2.723  & 14.67 \\
Ir    &  331  & 4.381  & 2.703  & 14.98 \\
Pt    &  144  & 4.526  & 2.704  & 15.99 \\
Au    &   62  & 4.748  & 2.815  & 18.32 \\
\end{tabular}
\end{ruledtabular}
\end{center}
\end{table}

\begin{table}[tbp]
\begin{center}
\caption{
Calculated energies,
lattice constants, and volumes
of the $\omega$ structure for the transition elements
in the magnetic states with nonzero magnetic moments.
The energies are relative to that of the NM FCC structure.
\label{tb:energies_elements_mag}
}
\begin{ruledtabular}
\begin{tabular}{lldddd}
 &                &
\multicolumn{1}{c}{Relative energy} &
\multicolumn{1}{c}{$a_\omega$} &
\multicolumn{1}{c}{$c_\omega$} &
\multicolumn{1}{c}{Volume} \\
 &                &
\multicolumn{1}{c}{(meV/atom)} &
\multicolumn{1}{c}{(\AA)} &
\multicolumn{1}{c}{(\AA)} &
\multicolumn{1}{c}{(\AA$^{3}$/atom)} \\
\hline
Fe & FM    &   33  & 4.177  & 2.375  & 11.96 \\
   & $++-$ &   35  & 3.992  & 2.386  & 10.97 \\
   & $+--$ &    1  & 3.970  & 2.397  & 10.91 \\
   & NM    &   91  & 3.850  & 2.418  & 10.35 \\
\\
Co & FM    & -120  & 3.953  & 2.434  & 10.98 \\
   & NM    &  166  & 3.899  & 2.390  & 10.49 \\
\\
Ni & FM    &   72  & 3.989  & 2.414  & 11.09 \\
   & NM    &  111  & 3.987  & 2.397  & 11.00 \\
\end{tabular}
\end{ruledtabular}
\end{center}
\end{table}

\begin{table}[htbp]
\begin{center}
\caption{
Calculated magnetic moments on atoms for the $\omega$ structure.
Note that the sites S2 and S3 are crystallographically equivalent
for the FM and the $+--$ states
but not for the $++-$ state.
The equivalent values are shown in parentheses.
\label{tb:magnetic_moments_elements}
}
\begin{ruledtabular}
\begin{tabular}{llddd}
 & 
 & 
\multicolumn{3}{c}{Magnetic moment ($\mu_B$)} \\
\cline{3-5}
&
&
\multicolumn{1}{C{1.5cm}}{S1} &
\multicolumn{1}{C{1.5cm}}{S2} &
\multicolumn{1}{C{1.5cm}}{S3} \\
\hline
Fe & FM    & 2.59  &  2.45  &  (2.45) \\
   & $++-$ & 1.56  &  1.72  &  -1.90  \\
   & $+--$ & 1.69  & -1.63  & (-1.63) \\
\\
Co & FM    & 1.60  &  1.67  &  (1.67) \\
\\
Ni & FM    & 0.62  &  0.66  &  (0.66) \\
\end{tabular}
\end{ruledtabular}
\end{center}
\end{table}

Table~\ref{tb:energies_elements_nm} shows calculated energies, lattice constants,
and volumes of the NM $\omega$ structure for 27 transition elements,
and Table~\ref{tb:energies_elements_mag} shows the values for the other magnetic states.
Table~\ref{tb:magnetic_moments_elements} gives calculated magnetic moments on atoms.
Volumes of the magnetic states with nonzero magnetic moments are larger than
that in the NM state.
%
The FM $\omega$ structure for Co and Ni are 285 and 39~meV/atom lower in energy
than the NM state, respectively.
For Fe,
the $+--$ magnetic state is the lowest in energy among the obtained magnetic states.
The FM $\omega$ Fe is 32 meV/atom higher in energy than the $+--$ magnetic state
and hence is thermodynamically less favorable.
The NM $\omega$ Fe,
which is 90 meV/atom higher in energy than the $+--$ state,
has the highest energy among the obtained magnetic states.

\begin{table*}[tbp]
\begin{center}
\caption{
Calculated energies of the transition elements
relative to that of the NM FCC structure in meV/atom.
\label{tb:energies_elements_all}
}
\begin{ruledtabular}
\begin{tabular}{ldddddddddd}
   &
\multicolumn{2}{c}{BCC} &
\multicolumn{2}{c}{FCC} &
\multicolumn{2}{c}{HCP} &
\multicolumn{4}{c}{$\omega$} \\
\cline{2-3}
\cline{4-5}
\cline{6-7}
\cline{8-11}
   &
\multicolumn{1}{C{1.2cm}}{NM} &
\multicolumn{1}{C{1.2cm}}{FM} &
\multicolumn{1}{C{1.2cm}}{NM} &
\multicolumn{1}{C{1.2cm}}{FM} &
\multicolumn{1}{C{1.2cm}}{NM} &
\multicolumn{1}{C{1.2cm}}{FM} &
\multicolumn{1}{C{1.2cm}}{NM} &
\multicolumn{1}{C{1.2cm}}{$+--$} &
\multicolumn{1}{C{1.2cm}}{$++-$} &
\multicolumn{1}{C{1.2cm}}{FM} \\
\hline
Sc & 56  &  & 0  &  & -49  &  & 8  &  &  & \\
Ti & 47  &  & 0  &  & -56  &  & -70  &  &  & \\
V  & -258  &  & 0  &  & 0  &  & -142  &  &  & \\
Cr & -397  &  & 0  &  & 9  &  & -66  &  &  & \\
Mn & 79  & 64  & 0  &  & -29  &  & 6  &  &  & \\
Fe & 314  & -170  & 0  & -20  & -80  &  & 91  & 1  & 35  & 33 \\
Co & 235  & -111  & 0  & -184  & 21  & -201  & 166  &  &  & -120 \\
Ni & 63  & 48  & 0  & -52  & 26  & -26  & 111  &  &  & 72 \\
Cu & 36  &  & 0  &  & 10  &  & 77  &  &  & \\
\\
Y  & 98  &  & 0  &  & -27  &  & 37  &  &  & \\
Zr & 46  &  & 0  &  & -40  &  & -40  &  &  & \\
Nb & -324  &  & 0  &  & -29  &  & -125  &  &  & \\
Mo & -431  &  & 0  &  & 14  &  & -35  &  &  & \\
Tc & 176  &  & 0  &  & -72  &  & -7  &  &  & \\
Ru & 513  &  & 0  &  & -116  &  & 137  &  &  & \\
Rh & 350  & 346  & 0  &  & 39  &  & 223  &  &  & \\
Pd & 45  &  & 0  &  & 31  &  & 107  &  &  & \\
Ag & 32  &  & 0  &  & 5  &  & 65  &  &  & \\
\\
Lu & 98  &  & 0  &  & -42  &  & 49  &  &  & \\
Hf & 107  &  & 0  &  & -74  &  & -34  &  &  & \\
Ta & -248  &  & 0  &  & 38  &  & -28  &  &  & \\
W  & -492  &  & 0  &  & 17  &  & 51  &  &  & \\
Re & 251  &  & 0  &  & -62  &  & 106  &  &  & \\
Os & 748  &  & 0  &  & -139  &  & 209  &  &  & \\
Ir & 633  & 626  & 0  &  & 74  &  & 331  &  &  & \\
Pt & 95  &  & 0  &  & 60  &  & 144  &  &  & \\
Au & 20  &  & 0  &  & 8  &  & 62  &  &  & \\
\end{tabular}
\end{ruledtabular}
\end{center}
\end{table*}

\begin{figure}[tbp]
\begin{center}
\includegraphics[width=\linewidth]{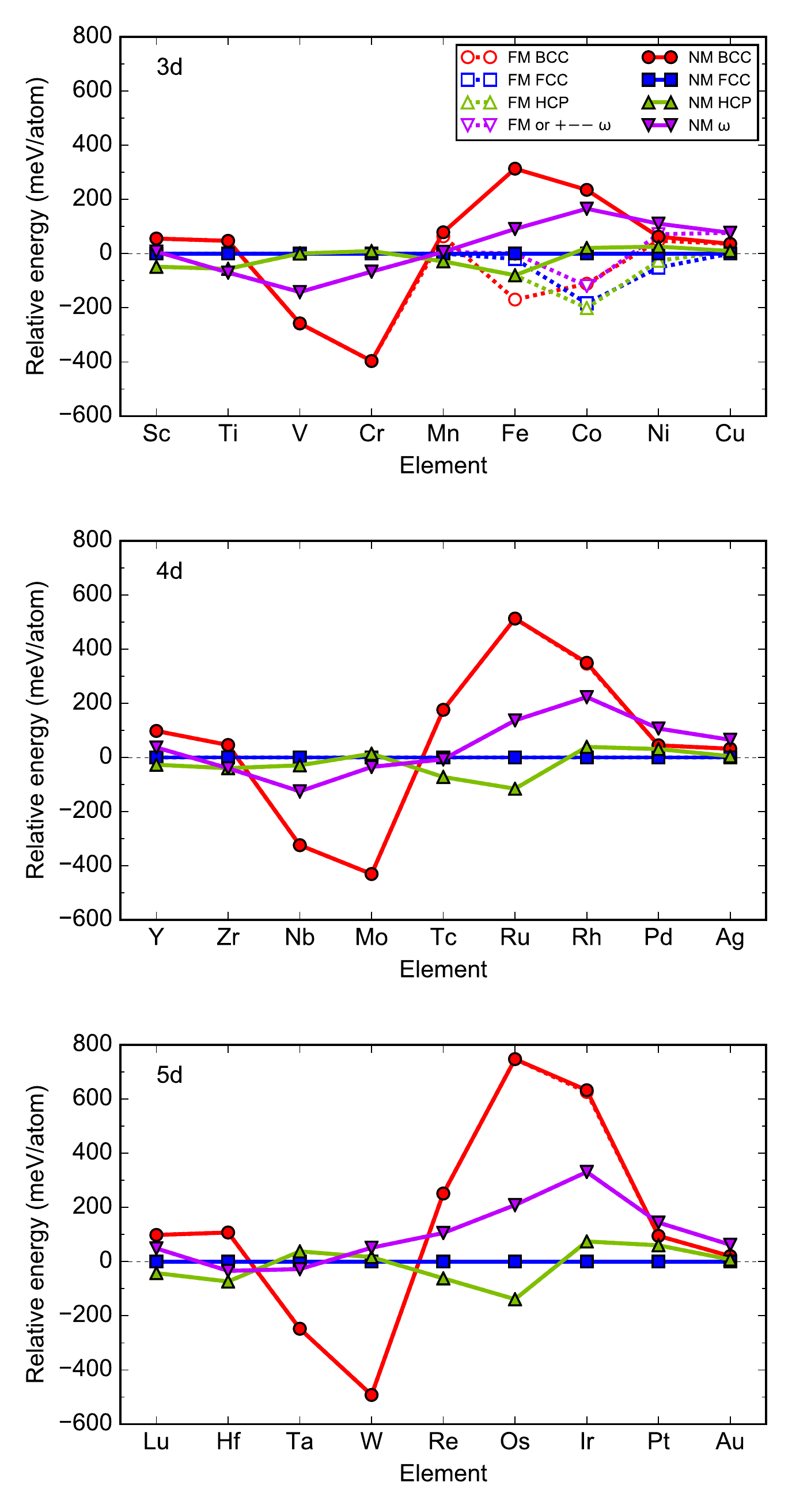}
\end{center}
\caption{
(Color online)
Calculated energies of the transition elements
relative to that of the NM FCC structure.
Red circle, blue square, green triangle, and purple inverse triangle symbols denote
the BCC, FCC, HCP, and $\omega$ structures, respectively.
Filled symbols connected by solid lines indicate the NM state,
while open symbols connected by dashed lines are for the magnetic state
that has the lowest energy among the investigated ones.
Note that the $+--$ $\omega$ structure is the lowest in energy only for Fe,
and the FM state is the lowest in energy for the other systems.
The lines are guides for the eyes.
\label{fig:energies_elements}
}
\end{figure}

Table~\ref{tb:energies_elements_all}
summarizes the energies relative to that of the NM FCC structure,
and Fig.~\ref{fig:energies_elements} visualizes the result.
For the NM state,
most of the elements in the same groups show the same energy sequences
for the investigated crystal structures.
For example,
the sequence for the group 8 elements (Fe, Ru, and Os) in the NM state is
HCP $\rightarrow$
FCC $\rightarrow$
$\omega$ $\rightarrow$
BCC
in order of increasing energy.
This result indicates that relative energies are roughly determined
from the occupation number for $d$ orbitals for these crystal structures.
This tendency has already been pointed out
for the BCC, FCC, and HCP structures
\cite{Skriver1985}.
The present calculations reveal that the $\omega$ structure also follows this rule.
The NM $\omega$ structure tends to be lower in energy than the NM FCC structure
for early transition elements except for those in the group 3 (Sc, Y, and Lu)
and to be higher for late transition elements.
This tendency is similar to that for the NM BCC structure.
The NM $\omega$ structure is also lower in energy than the NM BCC
for the elements in the groups 3, 4, 7, 8, and 9
and higher for the elements in the other groups.
For the elements in the groups 5 and 6 except for W,
the NM $\omega$ structure is the second lowest in energy
among the investigated crystal structures.

For the group 4 elements,
the relative energies of the $\omega$ structure are low
compared with the elements in the other groups.
The energies of the $\omega$ structure relative to those of the HCP,
which is observed in experiments at ambient temperature and pressure,
are $-13$, 0, and 39~meV/atom for Ti, Zr, and Hf, respectively.
These small energy differences indicate that
the $\omega$ structure is thermodynamically favorable for the group 4 elements.
The $\omega$ structure of the group 4 elements is actually observed in experiments
\cite{Jamieson1963, Xia1990, Pandey2014}.
Note that for Ti,
our computational result shows that the $\omega$ structure is lower in energy
than the HCP structure.
This result has also been shown in a previous report~\cite{Togo2013}
and hence is correct at least within DFT calculations using the GGA PBE functional.

The $+--$ $\omega$ Fe is 170~meV/atom higher in energy than the FM BCC Fe,
which is observed in experiments at ambient temperature and pressure.
This energy difference between the $\omega$ structure and the state in experiments
is much larger than those for the group 4 elements,
which implies that
the $\omega$ structure of Fe is thermodynamically more difficult to be formed
than that of the group 4 elements.
In contrast,
several experimental reports have claimed that
the $\omega$ structure can be formed in Fe-based alloys
~\cite{Ping2013, Liu2015}.
In these experiments,
the $\omega$ structure has been observed at twin boundaries of the BCC
or as precipitates in the BCC matrix.
Such kinds of structural imperfections and/or coherent stress at the interfaces
are maybe required to form the $\omega$ Fe.
Effect of solute elements is another possible reason
for the formation of the $\omega$ structure in the Fe-based alloys.
If the solute elements thermodynamically stabilize the $\omega$ structure
more than the BCC,
the Fe-based alloys may prefer to form the $\omega$ structure.
As mentioned above,
the NM $\omega$ structure is lower in energy than the NM BCC for the elements in the groups 3, 4, 7, 8, and 9,
and hence these elements are expected to stabilize the $\omega$ structure
more than the BCC.
In addition,
the FM $\omega$ Co is $-9$ meV/atom lower in energy than the FM BCC Co,
and hence Co is also expected to stabilize the $\omega$ structure
more than the BCC.

\begin{figure}[tbp]
\begin{center}
\includegraphics[width=\linewidth]{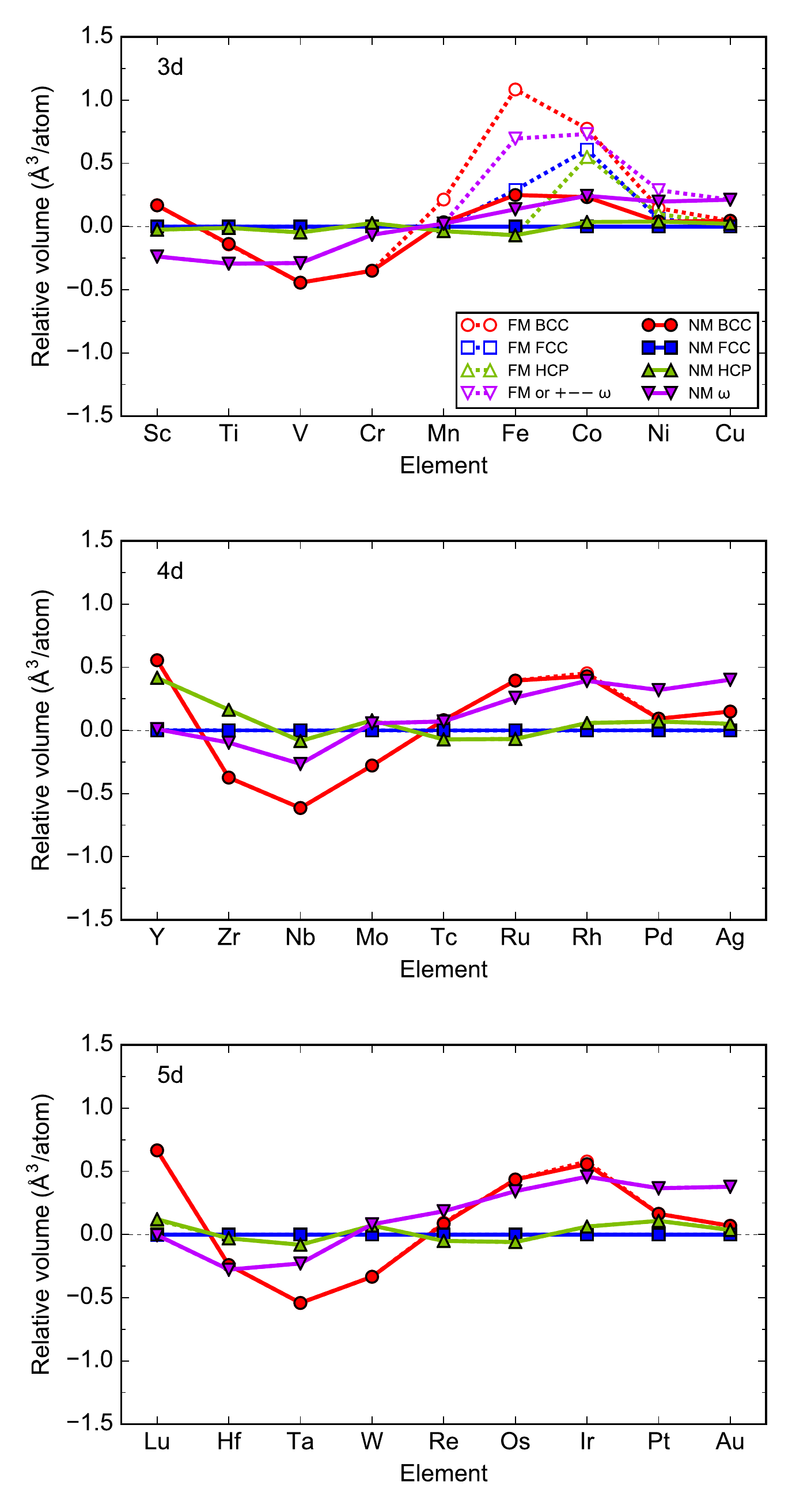}
\end{center}
\caption{
(Color online)
Calculated volumes of the transition elements
relative to that for the NM FCC structure.
Notations are the same as Fig.~\ref{fig:energies_elements}.
\label{fig:volumes_elements}
}
\end{figure}

Figure~\ref{fig:volumes_elements} summarizes the calculated volumes
relative to those of the NM FCC structure.
For the NM state,
most of the elements in the same groups show the same volume sequences
for the investigated crystal structure,
similarly to the case of the relative energies.
The NM $\omega$ structure tends to be smaller in volume than the NM FCC structure
for early transition elements and to be larger for late transition elements.


\begin{figure}[tbp]
\begin{center}
\includegraphics[width=\linewidth]{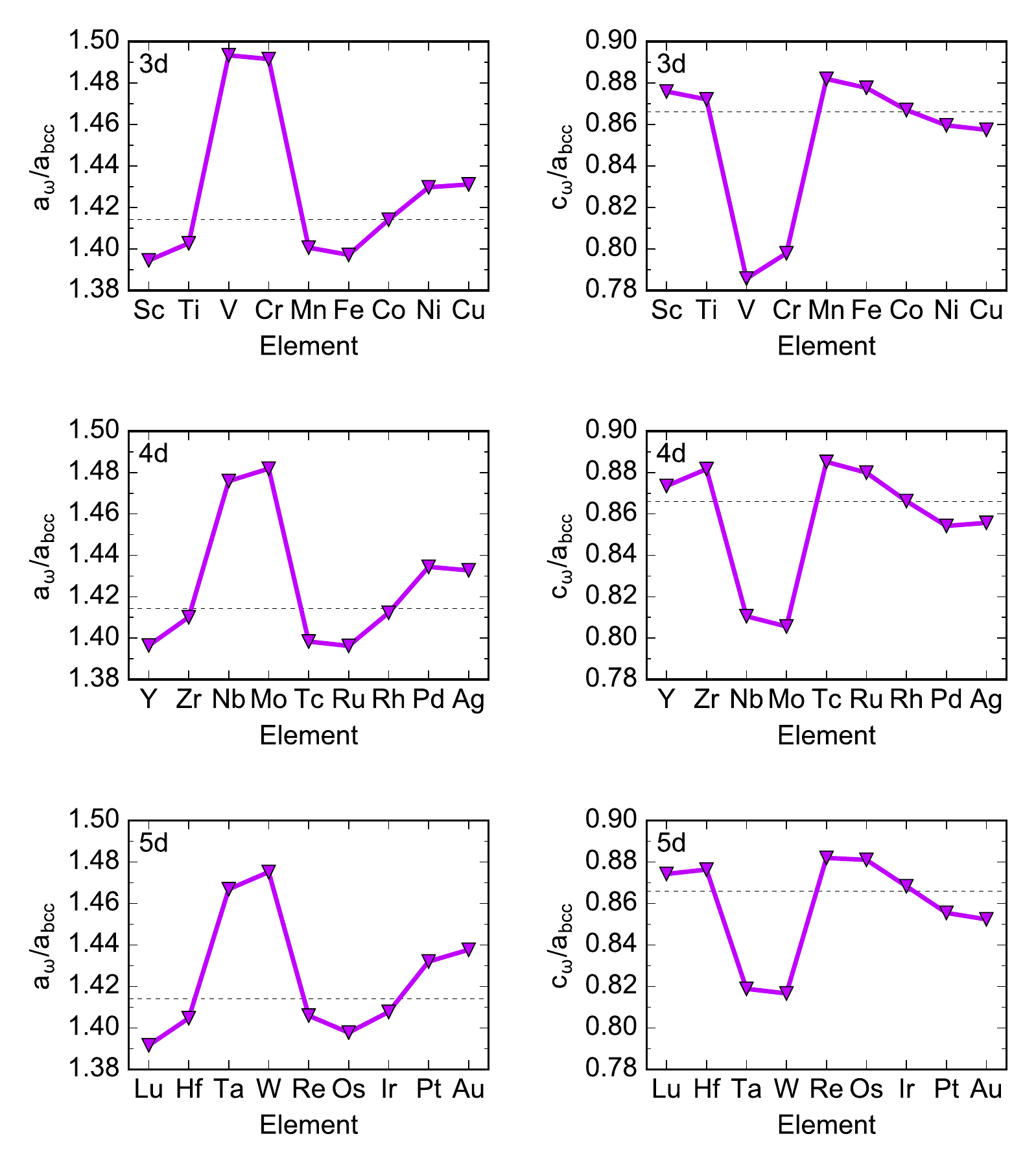}
\end{center}
\caption{
(Color online)
Ratios of calculated lattice constants of the NM $\omega$ structure
to that of the NM BCC structure.
The left and the right panels are for $a_\omega/a_\mathrm{BCC}$ and $c_\omega/a_\mathrm{BCC}$,
respectively.
Dashed horizontal lines indicate the values coherent with the BCC structure
described in Eqs.~(\ref{eq:a_omega}) and (\ref{eq:c_omega}).
Lines are guides for the eyes.
\label{fig:ratios_lattice_constants_elements}
}
\end{figure}

Figure~\ref{fig:ratios_lattice_constants_elements} shows the ratios of
calculated lattice constants of the NM $\omega$ structure
to that of the NM BCC structure.
Dashed holizontal lines indicate the values
coherent with the BCC structure
described in
Eqs.~(\ref{eq:a_omega}) and~(\ref{eq:c_omega}).
The values for the elements in the groups 5 and 6 are largely deviated
from the coherent ones
compared with the other transition elements.
Therefore,
the $\omega$ structure may be largely deformed for the elements in the groups 5 and 6
when it is coherent with the BCC structure.

\subsection{Mechanical stability of the $\omega$ structure}

\begin{figure*}[tbp]
\begin{center}
\includegraphics[width=\linewidth]{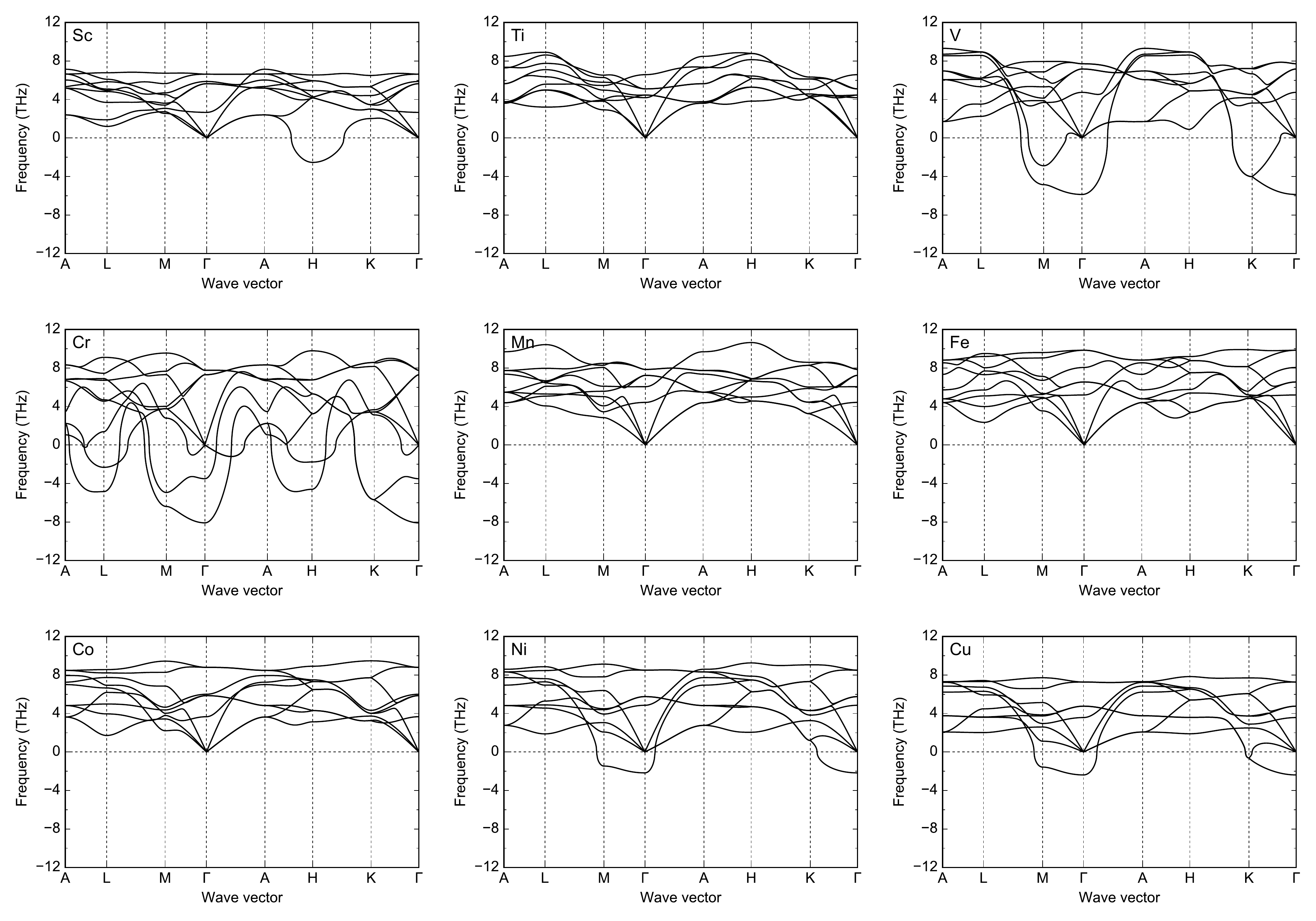}
\end{center}
\caption{
Calculated phonon dispersion relations of the NM $\omega$ structure
for the $3d$ transition elements.
Imaginary phonon frequencies are shown by negative values.
\label{fig:phonons_3d}
}
\end{figure*}

\begin{figure*}[tbp]
\begin{center}
\includegraphics[width=\linewidth]{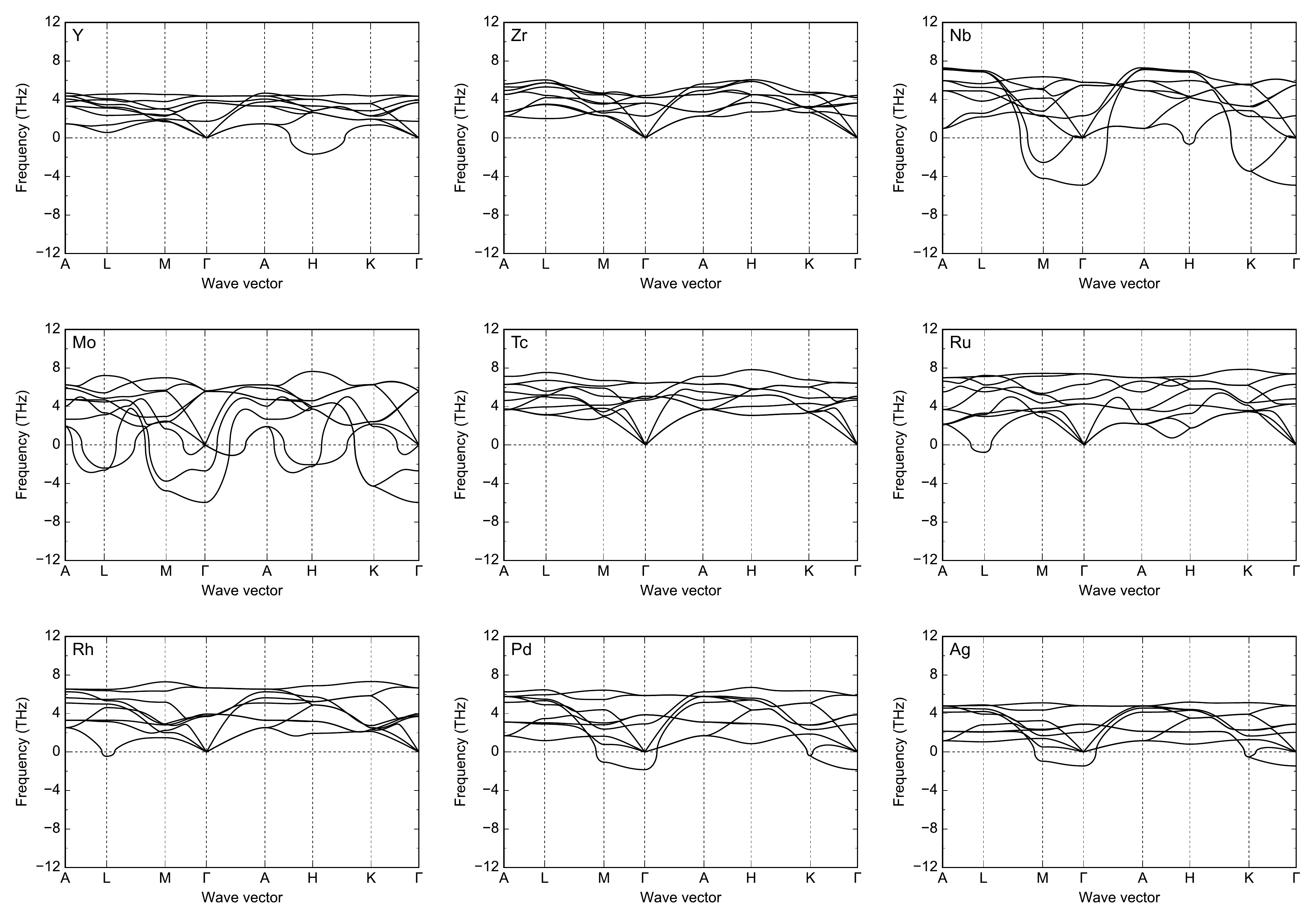}
\end{center}
\caption{
The same as Fig.~\ref{fig:phonons_3d} but for the $4d$ transition elements.
For Tc, which has no stable isotopes, we use the relative atomic mass of $^{99}$Tc, 98.906,
to calculate phonon dispersion relations.
Note that differences of the atomic mass only scale phonon frequencies.
\label{fig:phonons_4d}
}
\end{figure*}

\begin{figure*}[tbp]
\begin{center}
\includegraphics[width=\linewidth]{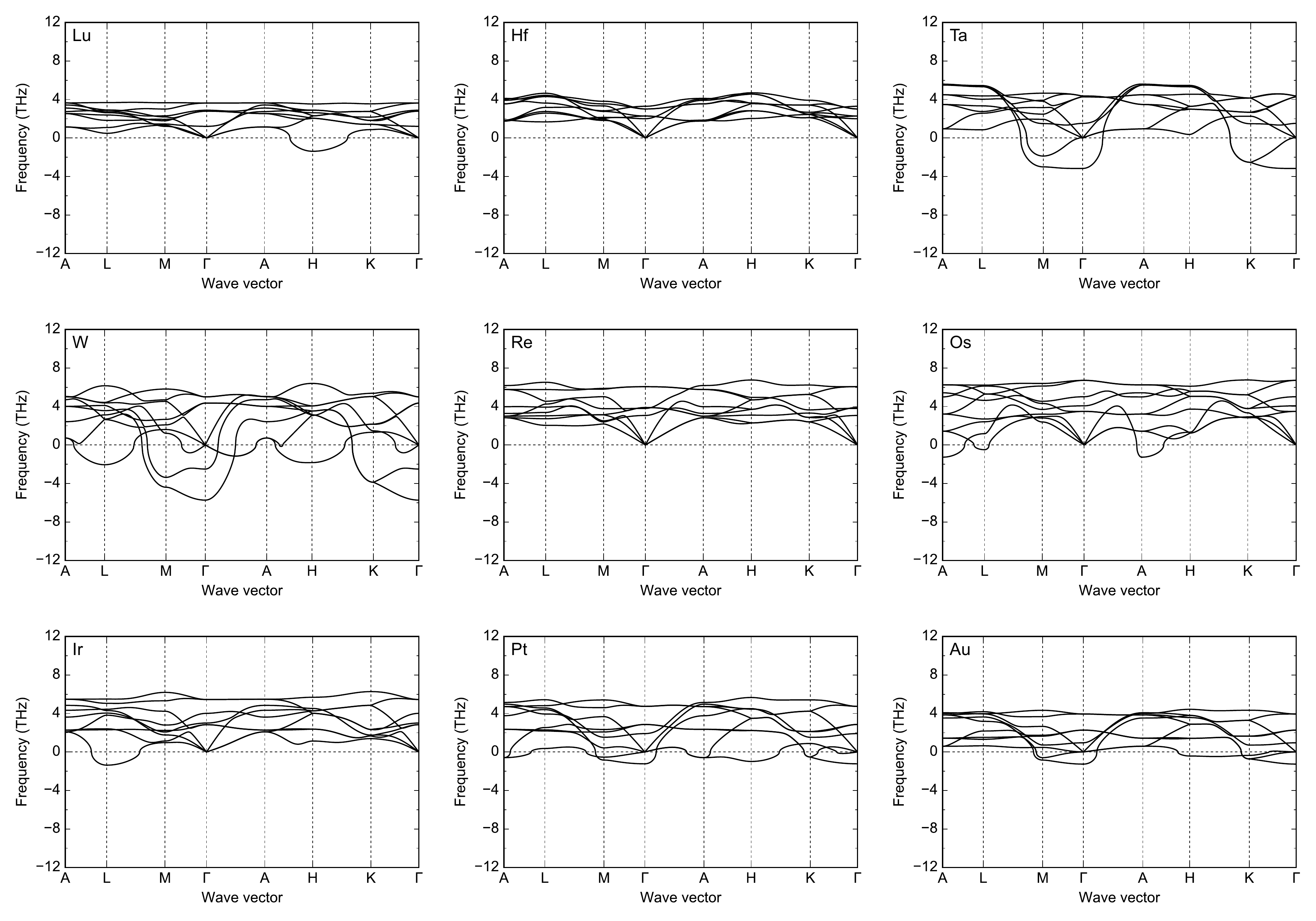}
\end{center}
\caption{
The same as Fig.~\ref{fig:phonons_3d} but for the $5d$ transition elements.
\label{fig:phonons_5d}
}
\end{figure*}

Figures~\ref{fig:phonons_3d}, \ref{fig:phonons_4d}, and \ref{fig:phonons_5d}
show calculated phonon dispersion relations of the NM $\omega$ structure
for the transition elements.
The elements in the same groups show similar shapes of the phonon dispersion relations.
Heavier elements in the same groups tend to have smaller absolute values of phonon frequencies.

The NM $\omega$ structure of Ti, Mn, Fe, Co, Zr, Tc, Hf, and Re
has no imaginary modes and hence is mechanically stable.
The $\omega$ structure of Ti and Zr is actually observed
at ambient temperature and pressure
\cite{Jamieson1963}.
Although Hf is included in the group 4 as well as Ti and Zr,
the $\omega$ Hf has not been observed in experiments
at ambient temperature and pressure
\cite{Xia1990, Pandey2014}.
The present calculations elucidate that
this is not because the $\omega$ Hf is mechanically unstable.

To our best knowledge,
the $\omega$ structure has not been reported for the group 7 elements (Mn, Tc, and Re),
while the $\omega$ structure for these elements is mechanically stable.
For the group 7 elements,
the energies of the $\omega$ structure relative to the HCP,
which is the lowest in energy among the investigated crystal structures,
are 35, 65, and 167 meV/atom for Mn, Tc, and Re, respectively.
These energy differences are higher than those for the group 4 elements.
In experiments, furthermore,
Mn forms the antiferromagnetic (AFM) A12 structure as the equilibrium state
at 4.2~K and at ambient pressure
\cite{Shull1953}.
First-principles calculations
suggest that
the energy of the AFM A12 structure is 60~meV/atom lower in energy than the NM HCP
\cite{Hafner2003}.
As a result,
we can estimate that
the $\omega$ Mn is 95~meV/atom higher in energy
than the experimental equilibrium state.
These results indicate that
although the $\omega$ structure of the group 7 elements
is mechanically stable,
they are thermodynamically less favorable than other crystal structures
compared with the group 4 elements.
Therefore,
the $\omega$ structure of the group 7 elements is probably more difficult
to be formed in experiments than that of the group 4 elements.

The $\omega$ structure of the elements in the groups 5 and 6 has
phonon modes with imaginary frequencies
and hence are mechanically unstable.
Several experimental reports, however,
have claimed that the $\omega$ structure can be formed in Ta and Mo
\cite{Hsiung2000, Cheng2013}.
In these experiments,
the $\omega$ structure has been observed in the matrix of the BCC structure,
and hence structural imperfections and/or coherent stress at the interfaces
between the $\omega$ and the BCC structures may affect
the formation of the $\omega$ structure.


\begin{figure*}[tbp]
\begin{center}
\includegraphics[width=0.65\linewidth]{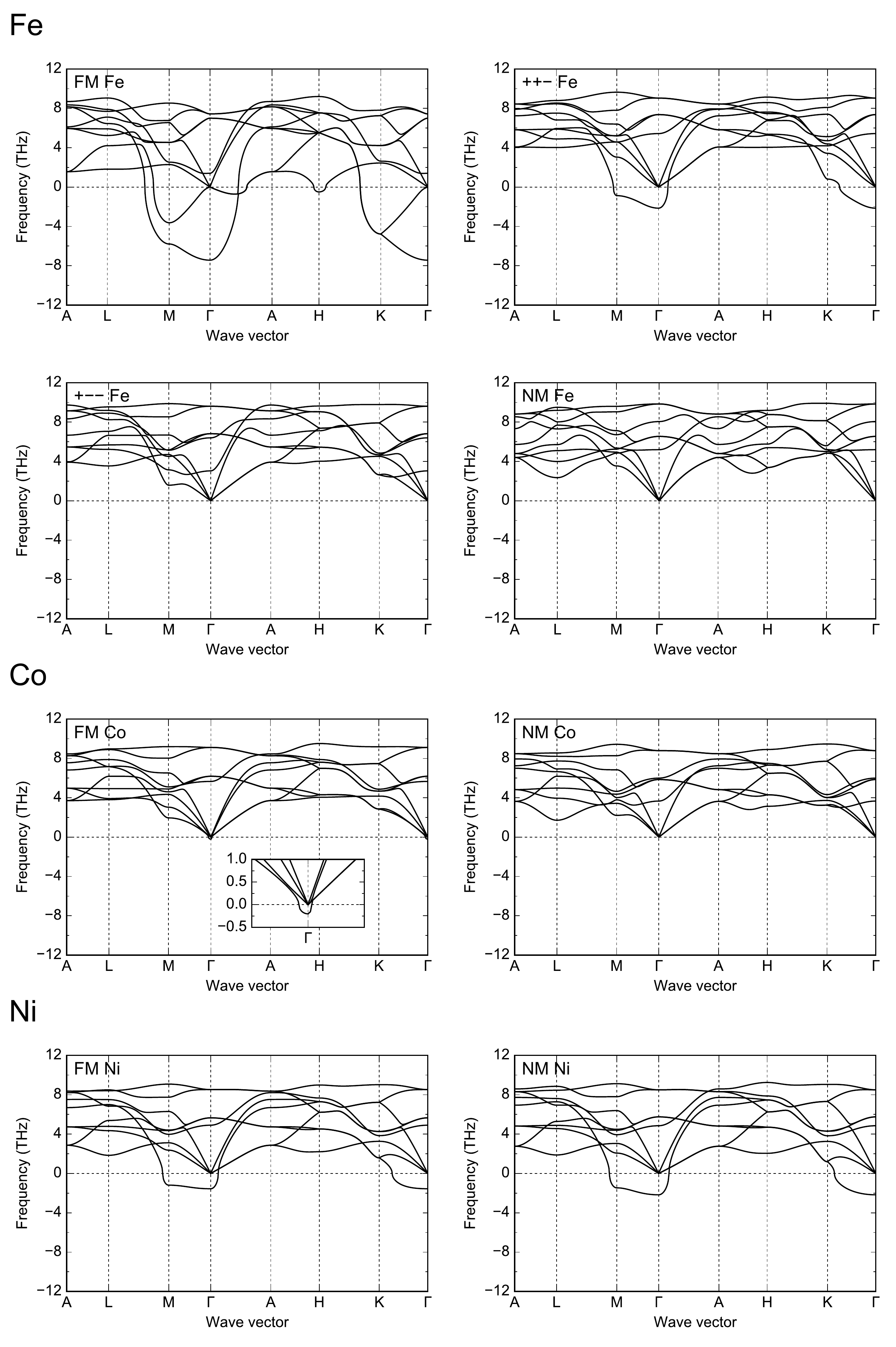}
\end{center}
\caption{
Calculated phonon dispersion relations of the $\omega$ structure in the obtained magnetic states for Fe, Co, and Ni.
The inset for the FM $\omega$ Co is the magnified view around the $\Gamma$ point,
which is shown to confirm a phonon mode with an imaginary frequency.
\label{fig:phonons_mag}
}
\end{figure*}

Figure~\ref{fig:phonons_mag} shows calculated phonon dispersion relations
of the $\omega$ structure in the obtained magnetic states for Fe, Co, and Ni.
Fe shows large differences among the magnetic states.
The FM and the $++-$ $\omega$ Fe have phonon modes with imaginary frequencies
and hence are mechanically unstable.
In contrast,
the $+--$ and the NM $\omega$ Fe have no phonon modes with imaginary frequencies
and hence are mechanically stable.
%
%
%
%
It has been well investigated that mechanical stability of the BCC and FCC Fe
largely depend on their magnetic states such as NM, FM, and PM ones
\cite{Hsueh2002, Kormann2012, Ikeda2014}.
The present result shows that magnetic states also affect the mechanical stability
of the $\omega$ Fe.
As shown above,
the $+--$ state is also the lowest in energy among the magnetic states for the $\omega$ structure,
and hence the $\omega$ Fe is expected to show the $+--$ state
if it is formed.

The FM $\omega$ Co has a phonon mode with an imaginary frequency at the $\Gamma$ point
(see the inset in Fig.~\ref{fig:phonons_mag}),
while the NM $\omega$ Co has no phonon modes with imaginary frequencies.
Since the absolute value of the imaginary phonon frequency is very small
for the FM $\omega$ Co,
It is difficult to conclude
whether the FM $\omega$ Co is actually mechanically unstable or not
at the present moment.
The $\omega$ Ni does not show qualitative differences between the FM and the NM states.

\subsection{Energetics for the $\omega$ structure of binary alloys}

The $\omega$ structure is frequently observed
in alloys based on the group 4 elements
\cite{Frost1954, Silcock1958, Sass1969, Hickman1969_TMSAIME, Cui2009, Wu2014, Ping2006,
Hatt1957, Hatt1960, Sass1969,
Jackson1970}.
It should be, therefore, interesting to investigate energetics of such alloys.

Since the occupation number for $d$ orbitals roughly determines
the relative energies of the crystal structures of our interest,
we can approximately estimate the energies of a binary alloy
A$_{1-x}$B$_{x}$
by linear interpolation as,
\begin{align}
E_{s}(\textrm{A}_{1-x}\textrm{B}_{x})
\approx
(1-x) E_{s}(\textrm{A})
+
x E_{s}(\textrm{B}),
\label{eq:energies_alloys}
\end{align}
where $E$ represents the energy per atom,
the symbols A and B represent the elements in the alloy,
$x$ represents the concentration of the element B,
and the subscript $s$ specifies the crystal structure.
Note that any kinds of atomic orderings are not considered in this equation.
The interpolated energy is actually the same as that of the phase-separation state
of the same crystal structure.
Nevertheless, this estimation is probably useful
to qualitatively discuss thermodynamical stability of crystal structures
for alloys with specific compositions.

\begin{figure*}[tbp]
\begin{center}
\includegraphics[width=\linewidth]{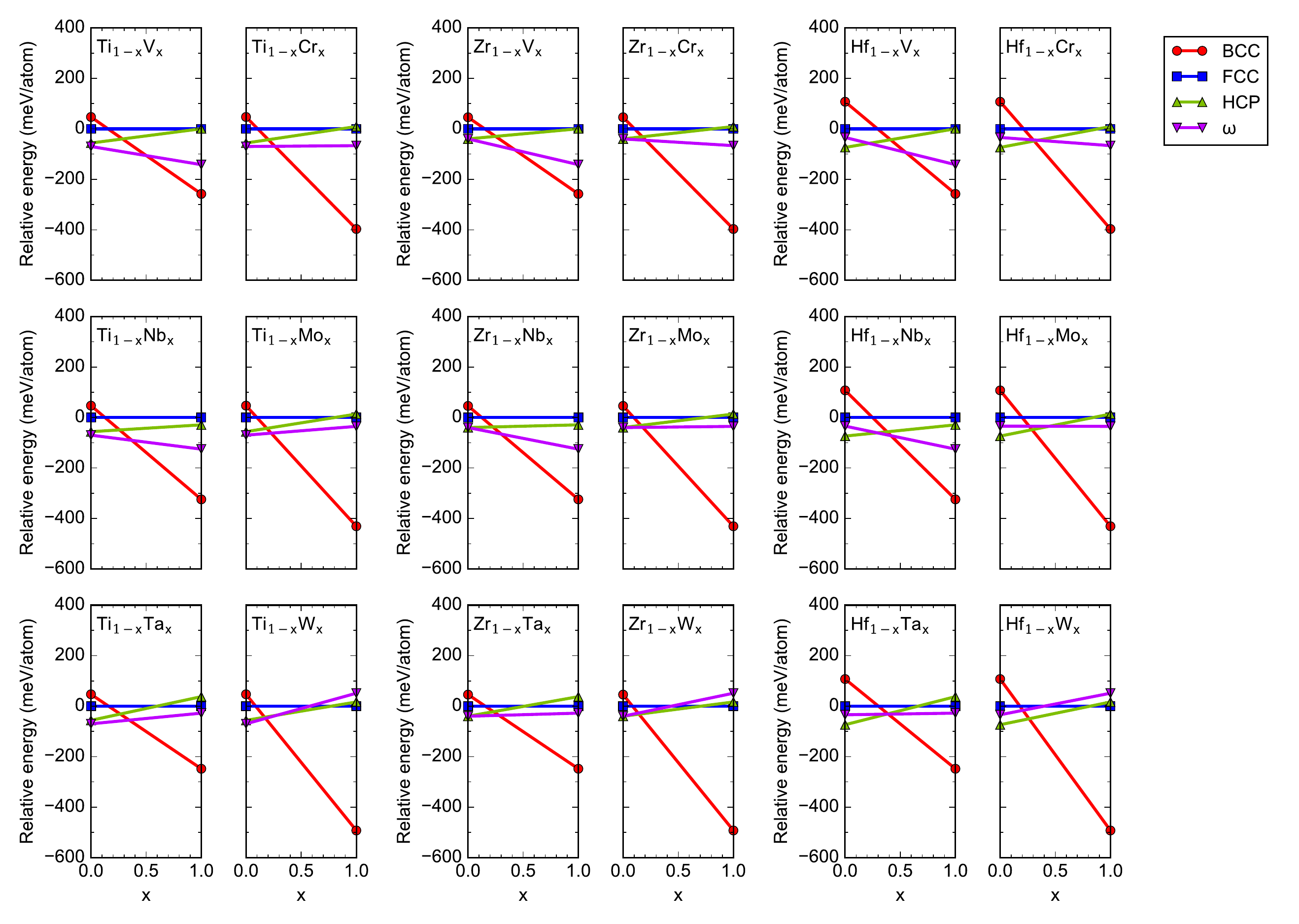}
\end{center}
\caption{
Estimated energies of the binary alloys composed of the elements in the group 4
and those in the groups 5 and 6.
The energies are relative to that of the NM FCC structure for each alloy.
The left side of each figure is for the elements in the group 4,
and the right side is for the elements in the groups 5 and 6.
\label{fig:energies_alloys}
}
\end{figure*}

Figure~\ref{fig:energies_alloys} shows
estimated energies of binary alloys composed of the elements in the group 4
and those in the groups 5 and 6.
Most of the binary alloys shown in the figure have the concentration range
where the $\omega$ structure is the lowest in energy
among the investigated crystal structures.
The group 4 elements are richer in these concentration ranges.
In experiments,
the $\omega$ structure is observed
for the alloys based on the group 4 elements with those in the groups 5 and 6
in the concentration ranges where the group 4 elements are richer
\cite{Frost1954, Hatt1957, Hatt1960, Silcock1958, Jackson1970, Ping2006, Cui2009, Hickman1969_TMSAIME}.
This experimental fact implies that
the $\omega$ structure is relatively favorable
for these alloys in the concentration ranges with rich group 4 elements.
The present computational result corresponds to this experimental fact.

The $\omega$ structure is favorable for these alloys
because of the following reason.
The HCP and the $\omega$ structures have almost the same energies for the group 4 elements,
while the HCP structure is substantially higher in energy
than the $\omega$ structure for the elements in the groups 5 and 6 except for W.
For the alloys composed of the group 4 elements and those in the groups 5 and 6, therefore,
the $\omega$ structure is estimated to be lower in energy than the HCP.
The $\omega$ structure is substantially lower in energy than the HCP 
only for the elements in the groups 5 and 6,
and hence this tendency is specific for the alloys
composed of the group 4 elements and those in the groups 5 and 6.

Note that even if the $\omega$ structure is the lowest in energy
at a specific concentration,
this does not mean that
the $\omega$ structure is the equilibrium state at this concentration.
When we consider a phase-separation state
composed of two different crystal structures with different compositions,
this phase-separation state can be lower in energy than the $\omega$ structure.
For example,
the estimated relative energy of the NM $\omega$ Ti$_{0.75}$V$_{0.25}$ to the NM FCC is $-88$ meV/atom,
while the mixture of 75 \% NM HCP Ti and 25 \% NM BCC V has
the relative energy of $-106$ meV/atom.
In experiments,
the $\omega$ structure in Ti-V alloys is actually thermodynamically metastable
and finally decomposes into the HCP and the BCC structures with different concentrations of
V atoms after sufficiently long-time aging 
\cite{Hickman1968}.

\section{CONCLUSION}

Properties of the $\omega$ structure are systematically investigated for 27 transition elements
from the viewpoints of thermodynamical and mechanical stability
using first-principles calculations.
The mechanical stability is investigated in terms of phonon frequencies.

The occupation number for $d$ orbitals roughly determines relative energy and volume
of the NM $\omega$ structure
as well as the other investigated crystal structures (BCC, FCC, and HCP).
For the group 4 elements (Ti, Zr, and Hf),
the $\omega$ structure is thermodynamically favorable
and is mechanically stable.
For the group 7 elements (Mn, Tc, and Re),
the $\omega$ structure is also mechanically stable,
but they are thermodynamically less favorable compared with the group 4 elements.
For the elements in the groups 5 and 6,
the lattice constants of the $\omega$ structure are largely deviated from the values
coherent with the BCC structure.

Several $3d$ transition elements, namely Fe, Co, and Ni,
have the $\omega$ structure with nonzero magnetic moments.
For the $\omega$ Fe,
the $+--$ state is the lowest in energy among the magnetic states
and is also mechanically stable.
Co and Ni show the FM $\omega$ structure,
and it is lower in energy than the NM state.

Energies of binary alloys composed of the elements in the group 4
and those in the groups 5 and 6 are estimated by linear interpolation.
Most of these alloys have a concentration range
where the $\omega$ structure is the lowest in energy
among the investigated crystal structures.
The group 4 elements are richer in these concentration ranges.
This result corresponds to the following experimental fact;
for the alloys composed of the group 4 and those in the groups 5 and 6,
the $\omega$ structure is observed in concentration ranges
where the group 4 elements are richer.


\begin{acknowledgments}

Funding by the Ministry of Education, Culture, Sports,
Science and Technology (MEXT), Japan, through Elements Strategy Initiative for
Structural Materials (ESISM) of Kyoto University,
is gratefully acknowledged.

\end{acknowledgments}

%

\end{document}